\documentclass[%
 reprint,
 amsmath,amssymb,
 aps,
 prd, 
]{revtex4-2}

\usepackage{graphicx}
\usepackage{dcolumn}
\usepackage{bm}
\usepackage{bbm}
\usepackage{amsthm}
\usepackage{aas_macros}
\usepackage{comment}
\usepackage{hyperref}
\usepackage{orcidlink}

\def\pd{{\rm d}}
\def\tr{{\rm Tr}}

\def\pa#1#2{\dfrac{\partial #1}{\partial #2}}
\def\pat#1#2{\dfrac{\partial^2 #1}{\partial {#2}^2}}

\def\tat#1#2{\dfrac{\pd^2 #1}{\pd {#2}^2}}

\def\deltadir{\delta_{\rm D}}

\def\expt{\mathbbm{E}}
\def\vari{\mathbbm{V}}
\def\cov{\mathbbm{Cov}}

\theoremstyle{definition}
\newtheorem{definition}{Definition}[section]

\theoremstyle{plain}

\theoremstyle{plain}
\newtheorem{proposition}{Proposition}[section]

\theoremstyle{plain}

\theoremstyle{remark}

\begin{document}

\preprint{APS/123-QED}

\title{A Geometric Theory of Cosmological Structure via Entropic Curvature in Wasserstein Space}

\author{Tsutomu T.\ Takeuchi\,\orcidlink{0000-0001-8416-7673}}
 \email{tsutomu.takeuchi.ttt@gmail.com}
\affiliation{%
 Division of Particle and Astrophysical Science, Nagoya University, \\
 Furo-cho, Chikusa-ku, Nagoya, 464–8602, Japan\\
}%
\affiliation{
 The Research Center for Statistical Machine Learning, the Institute of Statistical Mathematics, \\
 10-3 Midori-cho, Tachikawa, Tokyo 190–8562, Japan
}%

\date{\today}

\begin{abstract}
We construct a geometric framework for cosmological large-scale structure based on optimal transport theory and Wasserstein geometry. 
In this framework, Ricci curvature on the probability measure space $\mathcal{P}_2(M)$ is characterized by the geodesic convexity of entropy and is formulated as the response of probability distributions to optimal transport. 
We introduce effective Ricci curvatures $K_{\mathrm{eff}}^{(\infty)}$ and $K_{\mathrm{eff}}^{(N)}$ associated with Kullback--Leibler-type and R\'{e}nyi-type entropies, corresponding respectively to the curvature-dimension conditions CD$(K,\infty)$ and CD$(K,N)$. 
By localizing these curvatures to finite scales using local and reference measures, we construct curvature indicators applicable to observational data. 
Under a local quadratic approximation, the effective curvature reduces to the Hessian of the log-density, showing that conventional Hessian-based structure classifications arise as a limiting case of the present framework. 
We further show that effective curvature depends on observational scale and formulate this dependence as a scale flow, distinct from Ricci flow because it describes a change of resolution rather than a time evolution of geometry. 
Treating curvature as a random field then extends the statistical description of density fields: curvature statistics are given by higher-order weighted integrals of the power spectrum and by spatial derivatives of the correlation function, emphasizing geometric rather than amplitude information. 
This framework provides a unified connection between optimal transport geometry and cosmological structure analysis, and offers a new perspective on multiscale structure and nonlinear statistics.
\end{abstract}

\maketitle

\section{Introduction}
\label{sec:introduction}

Cosmological large-scale structure forms through the amplification of density fluctuations by gravitational instability and their nonlinear evolution, and exhibits characteristic geometric structures such as voids, filaments, and halos \citep[e.g.,][]{1980lssu.book.....P, fairall1998large_scale_structures, takeuchi2025physics}.
These structures are closely related to the statistical properties of the matter distribution as well as to the environmental dependence of galaxy formation and baryonic physics, and their quantification and classification remain central problems in cosmology.
Various theoretical and numerical methods have been proposed for this purpose \citep[e.g.,][]{2004RvMP...76.1211J, martinez2002statistics, takeuchi2025applications}.

Many conventional approaches are based on the local differential structure of the density or gravitational potential fields.
For example, cosmic web classifications using the Hessian of the density field or the tidal tensor identify voids, sheets, filaments, and halos through the sign structure of eigenvalues \citep{2007MNRAS.381...41H, 2009MNRAS.396.1815F, van_de_weygaert2011cosmic_web}.
Methods based on velocity fields and shear tensors are also widely used \citep{2012MNRAS.425.2049H, kitaura2026spectralhierarchycosmicweb}.
While these approaches characterize geometric structures, they do not explicitly incorporate global mass rearrangement or transport effects.
Similarly, statistical measures such as correlation functions and power spectra describe amplitude fluctuations but are primarily based on two-point statistics in Fourier space and do not directly capture spatial redistribution of matter \citep{1980lssu.book.....P, martinez2002statistics, takeuchi2025physics}.

In recent years, the Wasserstein distance from optimal transport theory has emerged as a powerful tool for comparing probability distributions and analyzing their geometry \citep{villani2009optimal, santambrogio2015optimal}.
In cosmology, it has been applied to reconstruct initial conditions from galaxy distributions and to Lagrangian-based structure analysis \citep[e.g.,][]{2002Natur.417..260F, 2003MNRAS.346..501B, 2024PhRvD.109l3512N, sartori2026backintime_voidfinder}.
Combined with topological data analysis, it has also been used to compare summaries from persistent homology \citep[e.g.,][]{edelsbrunner2010computational_topology, kerber2017geometry_persistence, 2023MNRAS.522.2697T}.
These studies show that Wasserstein distance effectively captures morphological differences between distributions \citep[e.g.,][]{Takeuchi2026OptimalTransport}.

However, existing approaches largely use the Wasserstein distance as a comparison metric without exploiting the intrinsic geometric structure of the probability measure space $\mathcal{P}_2(M)$, in particular its curvature.
In optimal transport theory, this space possesses a rich geometry in which Ricci curvature is defined via the geodesic convexity of entropy \citep{lott2009ricci,sturm2006geometryI,sturm2006geometryII}.
This formulation is closely related to Otto’s formal Riemannian structure \citep{Otto2001} and the gradient-flow formulation of the Jordan--Kinderlehrer--Otto scheme \citep{Jordan1998}, providing a unified description of diffusion and curvature.
Here curvature appears as the second variation of entropy and quantifies the geometric response of probability distributions to deformation.

Discrete analogues such as Ollivier-Ricci curvature define curvature via Wasserstein distances between local measures \citep{Ollivier2009RicciCurvature, ollivier2010ricci_survey, garcia_trillos_weber2024ollivier_ricci}.
While useful for network and discrete data analysis, they differ fundamentally from curvature defined on continuous measure spaces via entropy convexity and its second variation.
Morphological analysis that directly uses the continuous geometric structure of $\mathcal{P}_2(M)$ remains largely unexplored.

From this viewpoint, cosmological large-scale structure can be regarded not merely as a density field, but as the deformation of mass distributions within Wasserstein geometry \citep[e.g.,][]{Takeuchi2026OptimalTransport}.
Structural differences are then characterized not by static configurations but by their deformability, namely by curvature in measure space.
This provides a transport-geometric description of morphology distinct from conventional differential or statistical approaches.

In this work, we develop this perspective by localizing the effective Ricci curvature defined through entropy convexity and formulating it as a finite-scale observable.
We introduce two types of curvature based on Kullback--Leibler and R\'{e}nyi entropies, corresponding to infinite- and finite-dimensional geometric effects.
Using the directional decomposition of local effective curvature, we construct a classification framework based on mean curvature and anisotropy.
To our knowledge, no previous work has directly applied entropy-based Ricci curvature in Wasserstein geometry to the quantitative analysis of cosmological large-scale structure.
The present framework not only reinterprets existing Hessian-based classifications, but also provides a more general geometric structure that encompasses them.

The organization of this paper is as follows.
In Section~\ref{sec:optimal_transport_basics}, we review the foundations of optimal transport and measure-space geometry.
In Section~\ref{sec:effective_ricci_curvature}, we define the effective Ricci curvature based on entropy convexity.
In Section~\ref{sec:local_curvature}, we localize this curvature to finite scales and formulate it as an observable quantity.
In Section~\ref{sec:structure_classification}, we develop a classification of structures based on local effective curvature.
In Section~\ref{sec:estimator_structure}, we derive the corresponding estimator structure as a theoretical consequence.
In Section~\ref{sec:discussion_conclusion}, we discuss the implications of the framework and present our conclusions and outlook.
The appendices provide technical derivations, including the volume distortion coefficient $\beta_t^{K,N}$ in Appendix~\ref{app:beta_derivation}.

\section{Fundamentals of Optimal Transport and Geometry of Measure Spaces}
\label{sec:optimal_transport_basics}

\subsection{The Optimal Transport Problem: Monge and Kantorovich}

The optimal transport problem provides a framework for describing the rearrangement of probability distributions \citep{monge1781memoire, kantorovich1942translocation, villani2009optimal}.
Given probability measures $\mu,\nu\in\mathcal{P}_2(M)$, Monge’s formulation seeks a map $T:M\to M$ satisfying
\begin{align}
\nu = T_\#\mu
\end{align}
that minimizes
\begin{align}
\int c(x,T(x))\,\mu(\pd x).
\end{align}
Here $T_\#\mu$ denotes the pushforward measure defined by
\begin{align}
(T_\#\mu)(A)=\mu(T^{-1}(A)).
\end{align}

Kantorovich introduced a relaxed formulation based on a coupling $\pi\in\Pi(\mu,\nu)$:
\begin{align}
\inf_{\pi\in\Pi(\mu,\nu)}\int c(x,y)\,\pi(\pd x \,\pd y ).
\end{align}
This formulation allows mass splitting and guarantees existence of an optimal solution under general conditions.
In the quadratic cost case $c(x,y)=d(x,y)^2$, the optimal value defines the squared Wasserstein distance.
When $\mu$ is absolutely continuous, the optimal plan is induced by a transport map $T$.

\subsection{Probability Measures and the Wasserstein Distance}

Let $\mathcal{P}(M)$ denote the set of probability measures on a separable metric space $(M,d)$.
The subset with finite second moment is
\begin{align}
\mathcal{P}_2(M)
=
\left\{
\mu\in\mathcal{P}(M)
\;\middle|\;
\int d(x_0,x)^2\,\mu(\pd x )<\infty
\right\}.
\end{align}

For $\mu,\nu\in\mathcal{P}_2(M)$, the Wasserstein distance is defined by
\begin{align}
W_2(\mu,\nu)^2
=
\inf_{\pi\in\Pi(\mu,\nu)}
\int_{M\times M}
d(x,y)^2\,\pi(\pd x \,\pd y ).
\end{align}
Here $\pi$ satisfies
\begin{align}
\int_{M\times M} \pi(\pd x \,\pd y) & = 1,\\
\int_{M} \pi(\pd x ,B) &=\nu(B), \\
\int_{M} \pi(A,\pd y) &= \mu(A).
\end{align}

This distance represents the minimal cost required to rearrange $\mu$ into $\nu$ and defines a nontrivial geometric structure on $\mathcal{P}_2(M)$ \citep{villani2009optimal}.

\subsection{Wasserstein Geodesics and Displacement Interpolation}

$(\mathcal{P}_2(M),W_2)$ is a geodesic space.
For $\mu_0,\mu_1\in\mathcal{P}_2(M)$, there exists a geodesic $\{\mu_t\}_{t\in[0,1]}$ satisfying
\begin{align}
W_2(\mu_s,\mu_t)=|t-s|\,W_2(\mu_0,\mu_1).
\end{align}

When $M=\mathbb{R}^n$ and an optimal map $T$ exists,
\begin{align}
\mu_t
=
\bigl((1-t)\mathrm{Id}+tT\bigr)_\#\mu_0
\end{align}
defines the displacement interpolation.
On a Riemannian manifold,
\begin{align}
\mu_t = (F_t)_\#\mu_0,\qquad F_t(x)=\gamma_x(t)
\end{align}
with $\gamma_x$ geodesics connecting $x$ and $T(x)$.

These geodesics provide a natural notion of interpolation in measure space and play a central role in defining curvature through entropy convexity.

\subsection{Entropy and Its Gradient Flow}

For $\mu(\pd x)=\rho \pd x$, the Kullback--Leibler entropy is
\begin{align}
\mathcal{E}_{\infty}(\mu) = \int \rho \ln \rho\,\pd x.
\end{align}

Its Wasserstein gradient flow satisfies
\begin{align}
\pa{\mu_t}{t} = -\nabla_{W_2}\mathcal{E}_{\infty}(\mu_t).
\end{align}
Combining with the continuity equation
\begin{align}
\pa{\rho_t}{t} + \nabla\cdot(\rho_t v_t)=0,
\end{align}
and velocity field
\begin{align}
v_t = -\nabla \ln \rho_t,
\end{align}
one obtains
\begin{align}
\pa{\rho_t}{t} = \Delta \rho_t.
\end{align}
Thus the gradient flow of entropy coincides with the heat equation \citep{Jordan1998,Ambrosio2008GradientFlows}.

\subsection{Curvature-Dimension Condition and Entropy Convexity}

The curvature-dimension condition CD$(K,N)$ encodes a lower bound on Ricci curvature and an upper bound on dimension:
\begin{align}
\mathrm{Ric} \ge K,\qquad \dim \le N.
\end{align}

In metric measure spaces, curvature is defined via entropy convexity along Wasserstein geodesics \citep{lott2009ricci,sturm2006geometryI,sturm2006geometryII}.
For the KL entropy,
\begin{align}
\mathcal{E}_{\infty}(\mu)=\int \rho \ln \rho\,\pd x,
\end{align}
the CD$(K,\infty)$ condition is:

\begin{definition}[CD$(K,\infty)$]\label{def:CD_condition}
A space satisfies CD$(K,\infty)$ if for any $\mu_0,\mu_1\in\mathcal{P}_2(M)$ there exists a geodesic $\{\mu_t\}$ such that
\begin{align}
\mathcal{E}_{\infty}(\mu_t)
\le
(1-t)\mathcal{E}_{\infty}(\mu_0)
+
t\,\mathcal{E}_{\infty}(\mu_1)
-
\frac{K}{2}t(1-t)W_2(\mu_0,\mu_1)^2.
\end{align}
\end{definition}

This inequality quantifies curvature through entropy variation.
For finite $N$, the R\'{e}nyi entropy
\begin{align}
\mathcal{E}_{N}(\mu) = -\int \rho^{1-\frac{1}{N}}\,\pd x
\end{align}
leads to the CD$(K,N)$ condition involving the volume distortion coefficient $\beta_t^{K,N}$:
\begin{widetext}
\begin{align}
\rho_t^{-1/N}(\gamma_t)
\ge
(1-t)\beta_{1-t}^{K,N}(d(\gamma_0,\gamma_1))^{1/N}\rho_0^{-1/N}(\gamma_0)
+
t\,\beta_t^{K,N}(d(\gamma_0,\gamma_1))^{1/N}\rho_1^{-1/N}(\gamma_1).
\end{align}
\end{widetext}

These conditions provide a unified framework for describing curvature through entropy convexity and diffusion properties.
In what follows, we use this structure to define curvature on $\mathcal{P}_2(M)$.

\section{Definition of Effective Ricci Curvature Based on Entropy Convexity}
\label{sec:effective_ricci_curvature}

In the previous section, we introduced the geometric structure on the probability measure space $(\mathcal{P}_2(M),W_2)$, and organized Wasserstein geodesics and diffusion processes as gradient flows of entropy functionals.
Through this, we clarified that the time evolution of probability distributions is closely tied to geometry on the measure space.
In this section, we take this structure one step further, and define the curvature of the space from the geodesic convexity of entropy itself.
This formulation uses, in the reverse direction, the characterization of Ricci curvature in optimal transport theory \citep{lott2009ricci,sturm2006geometryI,sturm2006geometryII}, and provides a framework in which curvature is regarded as a deformation property of probability distributions.

In what follows, we use the KL-type entropy and the R\'{e}nyi-type entropy as two realizations of the same family, denoted by $\mathcal{E}_{\infty}(\mu)$ and $\mathcal{E}_{N}(\mu)$, respectively.
With this notation, it becomes clear that the diffusive geometry in the infinite-dimensional limit and the geometry accompanied by volume distortion in finite dimensions can be described in a unified manner under a single theory of entropy convexity.

\subsection{Wasserstein Geometry and KL-type Entropy}

We consider the probability measure space $(\mathcal{P}_2(M),W_2)$ introduced in the previous section.
Let $(M,d,m)$ be a complete separable metric measure space, and for $\mu=\rho m\in\mathcal{P}_2(M)$ define the KL-type entropy by
\begin{align}
\mathcal{E}_{\infty}(\mu)
=
\int_M \rho \ln\rho \,\pd m
\end{align}

When viewed along a Wasserstein geodesic $\{\mu_t\}_{t\in[0,1]}$ connecting $\mu_0,\mu_1\in\mathcal{P}_2(M)$, the change of entropy reflects not only the values of the density but also the geometry of mass rearrangement by optimal transport.
What expresses this point most clearly is Otto's formal Riemannian geometry.
From Otto's standpoint, $\mathcal{P}_2(M)$ is regarded as a formal infinite-dimensional Riemannian manifold, and geodesics are described by a velocity field $v_t$ satisfying the continuity equation
\begin{align}
\frac{\partial}{\partial t}\mu_t+\nabla\cdot(\mu_t v_t)=0
\end{align}
In particular, along the geodesic direction one can write $v_t=\nabla\phi_t$, and the norm of its tangent vector is given by
\begin{align}
\|\dot{\mu}_t\|_{W_2}^2
=
\int_M |v_t|^2 \,\pd\mu_t
\end{align}
Under this formal metric, the fact that the Hessian of the entropy functional $\mathcal{E}_{\infty}$ carries curvature information constitutes the core of the notion of curvature in measure space.

From this viewpoint, the drop of $\mathcal{E}_{\infty}$ at the midpoint along a Wasserstein geodesic
\begin{align}
(1-t)\mathcal{E}_{\infty}(\mu_0)
+
t\,\mathcal{E}_{\infty}(\mu_1)
-
\mathcal{E}_{\infty}(\mu_t)
\end{align}
is a natural quantity that measures the strength of the geodesic convexity of entropy.
By normalizing this by the transport distance, one can introduce a curvature quantity that is independent of scale.

\begin{definition}[KL-type effective Ricci curvature]
\label{def:effective_ricci_infty}
For $\mu_0,\mu_1\in\mathcal{P}_2(M)$ and a Wasserstein geodesic $\{\mu_t\}_{t\in[0,1]}$ connecting them,
\begin{widetext}
\begin{align}
K_{\mathrm{eff}}^{(\infty)}(\mu_0,\mu_1;t) =
\frac{2}{t(1-t)W_2^2(\mu_0,\mu_1)}
\left[
(1-t)\mathcal{E}_{\infty}(\mu_0)
+
t\,\mathcal{E}_{\infty}(\mu_1)
-
\mathcal{E}_{\infty}(\mu_t)
\right]
\end{align}
\end{widetext}
is defined.
\end{definition}
\medskip
\noindent
This definition captures Ricci curvature in Wasserstein geometry not as a tensor field, but as the entropy response to the geodesic deformation of probability distributions.
In particular, the larger $K_{\mathrm{eff}}^{(\infty)}$ is, the more strongly convex the entropy becomes along the geodesic, and the deformation of distributions is controlled more stably.

Furthermore, under the curvature-dimension condition {\rm CD}$(K,\infty)$,
\begin{align}
\mathcal{E}_{\infty}(\mu_t)
\le
(1-t)\mathcal{E}_{\infty}(\mu_0)
+
t\,\mathcal{E}_{\infty}(\mu_1)
-
\frac{K}{2}t(1-t)W_2^2(\mu_0,\mu_1)
\end{align}
holds, and therefore
\begin{align}
K_{\mathrm{eff}}^{(\infty)}(\mu_0,\mu_1;t)\ge K
\end{align}
follows.
Thus $K_{\mathrm{eff}}^{(\infty)}$ can be regarded as a quantity that extracts the curvature lower bound in the CD condition for each geodesic.

\subsection{Otto Structure, Second Variation, and Bochner-type Representation}

That the above definition is not merely a formal analogy can be confirmed through the second variation of $\mathcal{E}_{\infty}$.
Let $\{\mu_t\}$ be a sufficiently smooth Wasserstein geodesic, and write $\mu_t=\rho_t m$.
Using the continuity equation, one obtains
\begin{align}
\pa{\mathcal{E}_{\infty}(\mu_t)}{t}
=
\int_M \langle \nabla \ln\rho_t , v_t \rangle \,\pd\mu_t
\end{align}
Furthermore, following Otto's formal computation, differentiating once more with respect to $t$, and using the geodesic equation and integration by parts, one obtains a form
\begin{align}
\pat{\mathcal{E}_{\infty}(\mu_t)}{t}
=
\int_M
\left(
\|\nabla v_t\|^2 + \mathrm{Ric}(v_t,v_t)
\right)\pd\mu_t
\end{align}

This expression shows that the Hessian of entropy on Wasserstein space is decomposed into the sum of the deformation rate of the velocity field $\|\nabla v_t\|^2$ and the Ricci curvature term $\mathrm{Ric}(v_t,v_t)$.
That is, under the Otto metric,
\begin{align}
\mathrm{Hess}_{W_2}\mathcal{E}_{\infty}(\mu_t)[v_t,v_t]
=
\int_M
\left(
\|\nabla v_t\|^2 + \mathrm{Ric}(v_t,v_t)
\right)\pd\mu_t
\end{align}
is formally expressed.
In this sense, Ricci curvature appears directly in the second variation of entropy.
Therefore, if one writes, as the local limit at $t=0$,
\begin{align}
K_{\mathrm{eff}}^{(\infty)}(\mu)
=
\frac{1}{\|v_0\|_{L^2(\mu)}^2}
\pat{\mathcal{E}_{\infty}(\mu_t)}{t}\bigg|_{t=0}
\end{align}
this is a quantity obtained by normalizing the second variation of entropy in the geodesic direction, and gives the effective curvature in the geodesic direction.
This quantity is an indicator that measures the average effect of Ricci curvature along each tangent-vector direction.

Furthermore, this structure is closely connected with the Bochner inequality and the $\Gamma_2$ operator.
On a smooth Riemannian manifold, if one takes the generator to be $L=\Delta-\nabla V\cdot\nabla$, then in the notation of Bakry--\'Emery,
\begin{align}
\Gamma(f)
=
\frac{1}{2}\left[
L(f^2)-2fLf
\right]
=
|\nabla f|^2,
\end{align}
\begin{align}
\Gamma_2(f)
=
\frac{1}{2}
\left[
L\Gamma(f)-2\Gamma(f,Lf)
\right]
\end{align}
are defined.
In this case, the Bochner identity gives
\begin{align}
\Gamma_2(f)
=
\|\nabla^2 f\|^2
+
\mathrm{Ric}_V(\nabla f,\nabla f)
\end{align}
Here $\mathrm{Ric}_V=\mathrm{Ric}+\nabla^2V$ is the Bakry--\'{E}mery-type Ricci curvature.
Therefore, the inequality $\Gamma_2\ge K\Gamma$ simultaneously expresses the lower bound of Ricci curvature and the convexity control of the diffusion semigroup \citep{bakry2014analysis_markov, villani2009optimal}.

From the viewpoint of optimal transport, this is the same type of information as the statement that the second variation of entropy is controlled from below by $K$.
Therefore, the CD condition, the Bochner inequality, the $\Gamma_2$ structure, and the entropy convexity on Wasserstein space can be understood as the same notion of curvature expressed in different languages.

\subsection{Finite-dimensional Extension: R\'{e}nyi-type Entropy and CD$(K,N)$}\label{subsec:renyi_cdKN}

The KL-type entropy $\mathcal{E}_{\infty}$ corresponds to the infinite-dimensional limit.
In contrast, in order to incorporate a finite dimension $N$ explicitly, it is natural to introduce the R\'{e}nyi-type entropy
\begin{align}
\mathcal{E}_{N}(\mu)
=
-\int_M \rho^{1-\frac{1}{N}}\,\pd m
\end{align}
Here the notation $\mathcal{E}_{N}$ is adopted in order to make explicit that $\mathcal{E}_{\infty}$ and $\mathcal{E}_{N}$ are situated within the same theory of entropy convexity.
Through the power structure of the density, $\mathcal{E}_{N}$ is sensitive to the deformation of volume elements, and in particular reflects the contribution of high-density regions more strongly than the KL-type.
For this reason, $\mathcal{E}_{N}$ is suited not so much to the smoothing of diffusion itself, but rather to describing the geometry of volume compression and density concentration in finite-dimensional spaces.

The curvature-dimension condition {\rm CD}$(K,N)$ asserts that density deformation along a Wasserstein geodesic is controlled by the volume distortion coefficient $\beta_t^{K,N}$.
For a geodesic $\gamma_t$ matched by optimal transport,
\begin{align}
\rho_t^{-1/N}(\gamma_t)
&\ge
(1-t)\beta_{1-t}^{K,N}(r)^{1/N}\rho_0^{-1/N}(\gamma_0) \notag \\
&\qquad +
t\,\beta_t^{K,N}(r)^{1/N}\rho_1^{-1/N}(\gamma_1)
\end{align}
holds.
Here
\begin{align}
r=d(\gamma_0,\gamma_1)
\end{align}
is the distance between the pair of points matched by transport, and $t\in[0,1]$ is the geodesic parameter.
Therefore, $\beta_t^{K,N}(r)$ represents the distortion of the volume element at position $t$ along a geodesic of length $r$.
For example, in the case $K>0$,
\begin{align}
\beta_t^{K,N}(r)
=
\left[
\frac{\sin\!\bigl(t r\sqrt{K/(N-1)}\bigr)}
{t\,\sin\!\bigl(r\sqrt{K/(N-1)}\bigr)}
\right]^{N-1}
\end{align}
holds.
This coefficient is isomorphic to the volume deformation of Jacobi fields in comparison geometry, and the more positive the curvature is, the more the geodesic bundle converges and the more strongly the volume element is compressed.
Conversely, in negative curvature the geodesic bundle diverges, and volume expansion is promoted.

What rewrites this comparison-geometric information in the language of entropy is the geodesic convexity of $\mathcal{E}_{N}$.
Therefore, in complete parallel with the KL-type case, the finite-dimensional effective curvature can be defined by
\begin{widetext}
\begin{align}
K_{\mathrm{eff}}^{(N)}(\mu_0,\mu_1;t)
=
\frac{2}{t(1-t)W_2^2(\mu_0,\mu_1)}
\left[
(1-t)\mathcal{E}_{N}(\mu_0)
+
t\,\mathcal{E}_{N}(\mu_1)
-
\mathcal{E}_{N}(\mu_t)
\right]
\end{align}
\end{widetext}
This is the finite-dimensional version of $K_{\mathrm{eff}}^{(\infty)}$ based on $\mathcal{E}_{\infty}$, and is a quantity that expresses to what extent density concentration is controlled under a finite number of degrees of freedom.

\subsection{Entropy Convexity as a Unified Framework}

From the above discussion, it is seen that the KL-type and the R\'{e}nyi-type are not separate theories, but two aspects of a curvature theory based on entropy convexity.
$\mathcal{E}_{\infty}$ describes gradient flows and diffusion stability in the infinite-dimensional limit, whereas $\mathcal{E}_{N}$ describes the geometry of volume distortion and density concentration in finite dimensions.
More specifically,
\begin{align}
\mathcal{E}_{\infty}
&\longleftrightarrow
\begin{aligned}
&\text{heat flow, Otto metric,}\\
&\text{Bochner inequality, CD}(K,\infty)
\end{aligned}
\\
\mathcal{E}_{N}
&\longleftrightarrow
\begin{aligned}
&\text{volume distortion, comparison geometry,}\\
&\text{Jacobi fields, CD}(K,N)
\end{aligned}
\end{align}
hold as correspondences.
In the former, Ricci curvature appears in the Hessian of entropy, whereas in the latter curvature and dimension appear simultaneously through the volume distortion coefficient.

Therefore, the effective curvatures $K_{\mathrm{eff}}^{(\infty)}$ and $K_{\mathrm{eff}}^{(N)}$ are understood as quantities obtained by normalizing entropy convexity in the infinite-dimensional and finite-dimensional settings, respectively.
The advantage of this unified viewpoint lies in the fact that curvature is regarded not as a local tensor component, but as an \emph{effective geometric quantity} that collectively governs the deformation law of probability distributions, the stability of diffusion, the distortion of volume, and the dissipation of information.

Furthermore, this standpoint naturally connects to the issues of coarse-graining, scale dependence, and finite resolution that will be considered later.
This is because entropy convexity describes both the smoothing and the concentration of distributions, and therefore provides a foundation for tracking the transformation of geometric structure accompanying changes in observational resolution and coarse-graining radius.
In this sense, the effective Ricci curvature introduced in this section functions not merely as an abstract definition, but as a central concept that organizes scale-dependent geometric information in the subsequent discussion.

\section{Local Effective Curvature and Observables}
\label{sec:local_curvature}

Up to the previous chapter, based on the Wasserstein distance $W_2$ defined by optimal transport, we introduced the geometric structure on the probability measure space $(\mathcal{P}_2(M),W_2)$, and on that basis defined the effective Ricci curvatures $K_{\mathrm{eff}}^{(\infty)}$ and $K_{\mathrm{eff}}^{(N)}$ from the geodesic convexity of entropy.
These correspond to the second variation of entropy in Otto's formal Riemannian structure, and are quantities that give curvature as a geometric response to the deformation of probability distributions.
In this chapter, we localize this notion of curvature in Wasserstein geometry to finite regions and formulate it in a form corresponding to observable quantities.
That is, by evaluating entropy convexity along geodesics at finite scale, we construct curvature indicators applicable to real data.

\subsection{Local Effective Curvature}\label{subsec:local_effective_curvature}

Here locality is defined under coarse-graining depending on the spatial position $x$ and the scale $R$.
For a point $x\in M$ and a scale $R>0$, consider the geodesic ball with respect to the distance function $d$
\begin{align}
    B_R(x) =\{y\in M \mid d(x,y)<R\}
\end{align}
The measure restricted to this region is defined, by normalization, as a probability measure
\begin{align}
\mu_x^{(R)}(\pd y)
=
\frac{1}{\mu(B_R(x))}\,\mathbf{1}_{B_R(x)}(y)\,\mu(\pd y)
\end{align}
Furthermore, as a reference for comparing local structure, we introduce a reference measure $\nu_x^{(R)}$ on the same region.
Here we assume, for example, a uniform measure or a smoothed reference density.
From the optimal transport between these two measures, a Wasserstein geodesic $\{\mu_{x,t}^{(R)}\}_{t\in[0,1]}$ is determined.

At this point, in order to preserve symmetry and correspondence with the second variation, we evaluate local curvature using the drop of entropy at the midpoint $t=1/2$ of the geodesic.
\begin{widetext}
\begin{align}
K_{\mathrm{eff}}^{(\infty)}(x;R)
&=
-\frac{8}{W_2^2\!\left(\mu_x^{(R)},\nu_x^{(R)}\right)}
\left[
\mathcal{E}_{\infty}\!\left(\mu_{x,1/2}^{(R)}\right)
-
\frac{\mathcal{E}_{\infty}\!\left(\mu_x^{(R)}\right)+\mathcal{E}_{\infty}\!\left(\nu_x^{(R)}\right)}{2}
\right], \\
K_{\mathrm{eff}}^{(N)}(x;R)
&=
-\frac{8}{W_2^2\!\left(\mu_x^{(R)},\nu_x^{(R)}\right)}
\left[
\mathcal{E}_{N}\!\left(\mu_{x,1/2}^{(R)}\right)
-
\frac{\mathcal{E}_{N}\!\left(\mu_x^{(R)}\right)+\mathcal{E}_{N}\!\left(\nu_x^{(R)}\right)}{2}
\right]
\end{align}
\end{widetext}
Here $\mathcal{E}_{\infty}$ and $\mathcal{E}_{N}$ denote the KL-type and R\'{e}nyi-type entropies, respectively (cf.\ Section~\ref{sec:effective_ricci_curvature}).
This definition normalizes the second variation of entropy along the geodesic direction by the transport distance $W_2^2$, and thus provides a quantity that locally evaluates curvature in Wasserstein geometry.

\subsection{Theoretical Meaning and Sensitivity of Curvature}

$\mathcal{E}_{\infty}$ corresponds to the CD$(K,\infty)$ condition, and $\mathcal{E}_{N}$ corresponds to the CD$(K,N)$ condition.
Therefore
\begin{itemize}
    \item $K_{\mathrm{eff}}^{(\infty)}$: curvature in the infinite-dimensional limit,
    \item $K_{\mathrm{eff}}^{(N)}$: curvature including finite-dimensional effects
\end{itemize}
are understood accordingly.
Since the logarithm appears in the KL-type, it is sensitive to flattening including low-density regions, whereas in the R\'{e}nyi-type powers of the density appear, so that contributions from high-density regions are emphasized.
Therefore
\begin{itemize}
    \item $K_{\mathrm{eff}}^{(\infty)}$: sensitive to diffusion,
    \item $K_{\mathrm{eff}}^{(N)}$: sensitive to concentrated structures
\end{itemize}
this division of roles is established.

\subsection{Indicators as Observables}

Define the difference and ratio of the two curvatures as
\begin{align}
    \Delta K^{(N)}(x;R) &\equiv K_{\mathrm{eff}}^{(N)}(x;R)-K_{\mathrm{eff}}^{(\infty)}(x;R),\\
    \mathcal{R}^{(N)}(x;R) &= \frac{K_{\mathrm{eff}}^{(N)}(x;R)}{K_{\mathrm{eff}}^{(\infty)}(x;R)}
\end{align}

$\Delta K^{(N)}$ represents the increment of the contribution of concentrated structures, and $\mathcal{R}^{(N)}$ represents the relative balance between homogenization and concentration.
Therefore, local curvature is decomposed not as a single quantity but into a \emph{diffusive component} and a \emph{concentrated-structure component}.
By these two components, it becomes possible to characterize the geometric structure of the density field in a multifaceted manner.
The indicators defined in this chapter have a form that can be evaluated for finite samples, and are naturally extended to geometric analysis of discrete point distributions and to the quantification of scale-dependent structures.

\section{Classification of Cosmological Large-Scale Structure by Local Effective Curvature}
\label{sec:structure_classification}

This characterization arises as a local limit of the transport-geometric curvature defined through entropy convexity, and is not assumed \emph{a priori} as in conventional Hessian-based classification schemes.
The local effective curvatures $K_{\mathrm{eff}}^{(\infty)}(x;R), K_{\mathrm{eff}}^{(N)}(x;R)$ introduced in the previous section, together with their difference $\Delta K^{(N)}(x;R)$ and ratio $\mathcal{R}^{(N)}(x;R)$, are quantities that evaluate the geodesic convexity of entropy in Wasserstein geometry at finite scale, and are indicators that locally extract the geometric structure on the probability measure space $\mathcal{P}_2(M)$.
In this section, we provide a framework for classifying the geometric structure of cosmological density fields using these quantities.
In particular, $\Delta K^{(N)}$ represents the absolute incremental sensitivity to high-density structures, and $\mathcal{R}^{(N)}$ functions as a relative indicator that expresses to what extent this increment dominates relative to the KL-type curvature.

\subsection{Basic Principles and Simple Local Models}

Cosmological density fields are, to first approximation, classified into voids, filaments, and halos.
In this study, we regard these as the ease of deformation of local probability distributions, and characterize them by the behavior of local effective curvature.
For each structure, we approximate the local density distribution in a form in which geometric characteristics appear explicitly.

In voids, the density is low at the center and increases toward the surroundings.
As a local model,
\begin{align}
\rho(x)\simeq \rho_0\left(1+\epsilon |x|^2\right), \qquad \epsilon>0
\end{align}
is adopted.
This is a distribution in which the density is minimal at the origin and increases isotropically.

Since halos have strongly concentrated structures,
\begin{align}
\rho(x)\simeq \rho_0 \exp\left(-\frac{|x|^2}{\sigma^2}\right)
\end{align}
is used as an approximation.
This distribution contracts isotropically, and the density gradient is steep toward the center.

Filaments have anisotropic structures elongated in one direction.
Decomposing the coordinates as
\begin{align}
x=(x_\parallel,x_\perp)
\end{align}
one can approximate
\begin{align}
\rho(x)\simeq \rho_0\exp\left(-\frac{|x_\perp|^2}{\sigma_\perp^2}\right)
\end{align}
This is a structure that is nearly uniform along the filament direction and has contraction only in the perpendicular direction.

\subsection{Local Quadratic Approximation and Wasserstein Curvature}\label{subsec:local_quadratic_wasserstein}

\begin{figure}[t]
\centering
\includegraphics[width=\linewidth]{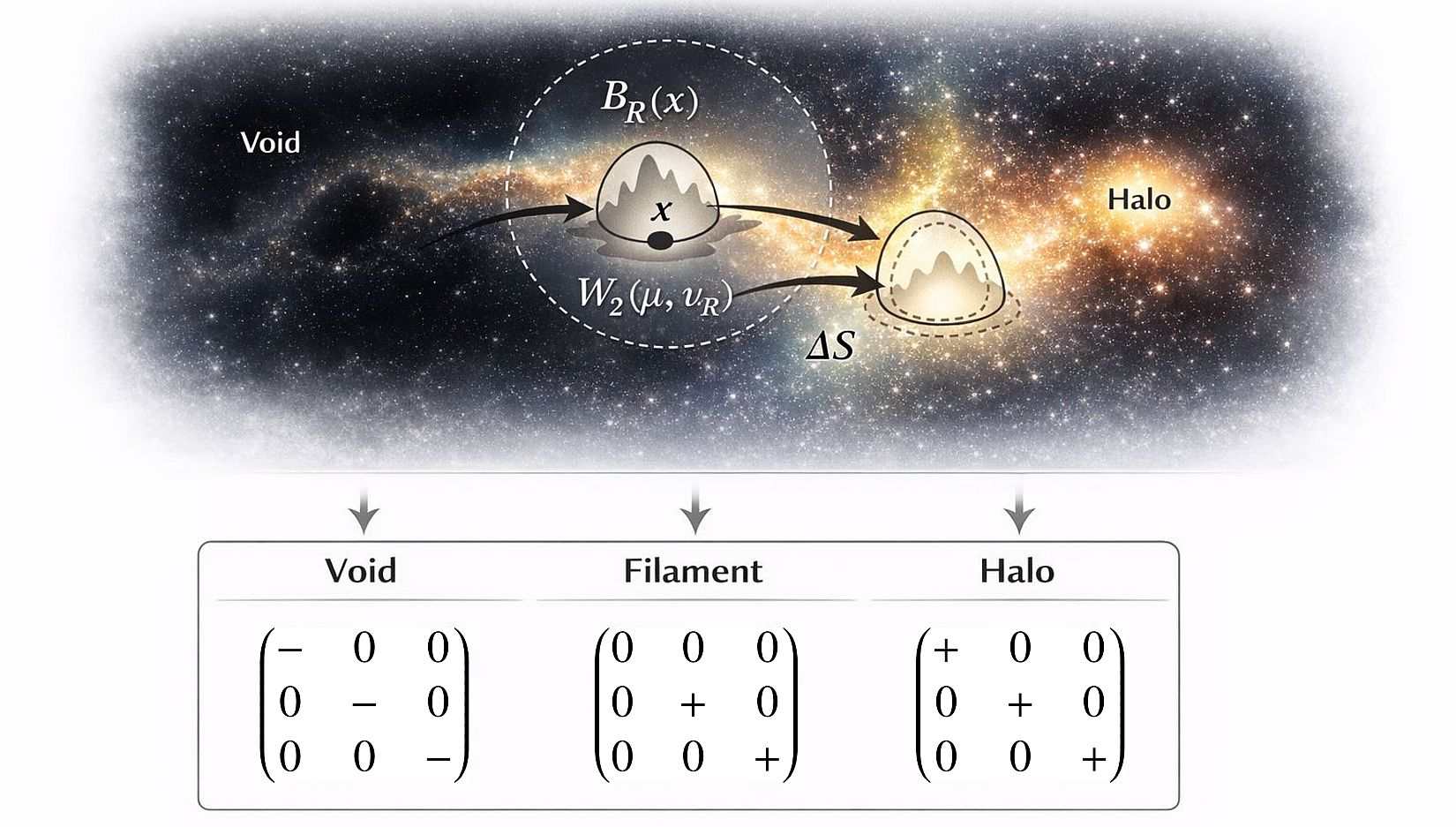}
\caption{
Schematic illustration of the local effective curvature in large-scale structure.
Top: a local region $B_R(x)$ defines a probability measure $\mu_x^{(R)}$, which is compared to a reference measure $\nu_x^{(R)}$ through optimal transport.
The curvature is quantified by the entropy variation $\Delta S$ along the Wasserstein geodesic.
Void regions correspond to low-density, dark areas, while halo regions correspond to high-density, bright concentrations.
Bottom: the corresponding local curvature structure is characterized by the Hessian of the log-density $H_x=-\nabla^2 \ln \rho(x)$.
Different eigenvalue configurations distinguish void, filament, and halo structures.
}
\label{fig:local_curvature}
\end{figure}

Expand the density distribution locally as
\begin{align}
\ln \rho(y)
=
\ln \rho(x)
-
\frac{1}{2}(y-x)^{\top}H_x(y-x)
+
O(|y-x|^3)
\end{align}
Here
\begin{align}
H_x=-\nabla^2 \ln \rho(x)
\end{align}
is the curvature matrix of the log-density, and its eigenvalue structure governs the local shape.

For the local measure $\mu_x^{(R)}$, using the affine deformation
\begin{align}
T_{\varepsilon,A}(y)=x+(I+\varepsilon A)(y-x)
\end{align}
define the reference measure
\begin{align}
\nu_x^{(R)}=(T_{\varepsilon,A})_{\#}\mu_x^{(R)}
\end{align}
In this case, using the local covariance
\begin{align}
\Sigma_x^{(R)}
=
\int_{B_R(x)} (y-x)(y-x)^{\top}\,\pd\mu_x^{(R)}(y)
\end{align}
one obtains
\begin{align}
W_2^2\!\left(\mu_x^{(R)},\nu_x^{(R)}\right)
=
\varepsilon^2 \tr(A\Sigma_x^{(R)}A^{\top})
+
O(\varepsilon^3)
\end{align}
Expanding the midpoint drop of entropy yields
\begin{align}
K_{\mathrm{eff}}^{(\infty)}(x;R;A)
=
\frac{\tr(H_x A\Sigma_x^{(R)}A^{\top})}
{\tr(A\Sigma_x^{(R)}A^{\top})}
+
O(R+\varepsilon)
\end{align}
Therefore, $K_{\mathrm{eff}}^{(\infty)}$ is interpreted as the averaged eigenvalue of the log-density curvature $H_x$ along the deformation direction $A$ in Wasserstein geometry.

Here this local expression is understood as the local limit of the second variation of entropy in the Otto formulation introduced in Section~3.2.
That is, since the entropy Hessian is governed by the second derivative of the log-density for a local linear deformation $v(y)=A(y-x)$, the local effective curvature is a quantity in which the finite-scale evaluation of Ricci curvature on measure space degenerates locally to $H_x$.

Similarly, for the R\'{e}nyi type,
\begin{align}
\Sigma_{N,x}^{(R)}
=
\dfrac{\displaystyle 
\int_{B_R(x)} (y-x)(y-x)^{\top}\rho(y)^{1-\frac{1}{N}}\,\pd y
}{\displaystyle 
\int_{B_R(x)} \rho(y)^{1-\frac{1}{N}}\,\pd y
}
\end{align}
using this, one can write
\begin{align}
K_{\mathrm{eff}}^{(N)}(x;R;A)
=
c_N(x;R)
\dfrac{\displaystyle \tr(H_x A\Sigma_{N,x}^{(R)}A^{\top})}
{\displaystyle\tr(A\Sigma_x^{(R)}A^{\top})}
+
O(R+\varepsilon)
\end{align}
This is a curvature with weight on high-density regions.

In particular, if one chooses $A=\mathbf{e}\mathbf{e}^{\top}$ representing deformation along a single direction $\mathbf{e}$, the directional curvature
\begin{align}
K_{\mathrm{eff}}^{(\infty)}(x;R;\mathbf{e})
\equiv
K_{\mathrm{eff}}^{(\infty)}(x;R;A=\mathbf{e}\mathbf{e}^{\top})
\end{align}
is defined, and in the isotropic limit one obtains
\begin{align}
K_{\mathrm{eff}}^{(\infty)}(x;R;\mathbf{e})
\simeq \mathbf{e}^{\top}H_x\mathbf{e}
\end{align}
Similarly, $K_{\mathrm{eff}}^{(N)}(x;R;\mathbf{e})$ is also defined.

\subsection{Characteristics of Each Structure}

In the local quadratic approximation, the curvature matrix is written as
\begin{align}
H_x = \mathrm{diag}(\lambda_1,\lambda_2,\lambda_3).
\end{align}
The eigenvalue signature provides a complete local classification of geometric structure:
\begin{align}
\text{Void:} \quad & \lambda_1,\lambda_2,\lambda_3<0, \notag \\
\text{Filament:} \quad & \lambda_1\simeq0,\ \lambda_2,\lambda_3>0, \\
\text{Halo:} \quad & \lambda_1,\lambda_2,\lambda_3>0. \notag
\end{align}
From this representation, the characteristics of each structure follow directly.
\medskip

\noindent
{\bf Void: }~
Since the density is minimal at the center,
\begin{align}
   \nabla^2 \ln \rho(x) > 0\quad \Rightarrow \quad H_x < 0,
\end{align}
and all eigenvalues are negative.
Therefore, curvature is weak in all directions and nearly isotropic, and the directional dependence is small.
Moreover, since high-density concentration is weak, $\Delta K^{(N)}$ is also small, and $\mathcal{R}^{(N)}$ is close to unity, or cannot be regarded as a stable indicator when $K_{\mathrm{eff}}^{(\infty)}$ is sufficiently small.
\medskip

\noindent
{\bf Halo:}~ 
For a centrally concentrated structure,
\begin{align}
    H_x=\frac{2}{\sigma^2}I,
\end{align}
and all eigenvalues are positive and large.
Therefore, the curvature is strong in all directions and isotropic.
Furthermore, due to central concentration, the R\'{e}nyi-type curvature tends to respond more strongly than the KL-type curvature, and
\begin{align}
    \Delta K^{(N)}(x;R) > 0, \qquad \mathcal{R}^{(N)}(x;R) > 1
\end{align}
becomes pronounced.
In particular, $\Delta K^{(N)}$ represents the absolute strength of concentrated structures, and $\mathcal{R}^{(N)}$ represents the relative importance of that concentration effect within the total curvature.
\medskip

\noindent
{\bf Filament:}~  
For anisotropic structures,
\begin{align}
    H_x=\mathrm{diag}\!\left(0,\frac{2}{\sigma_\perp^2},\frac{2}{\sigma_\perp^2}\right),
\end{align}
and the curvature differs significantly depending on direction.
Along the filament direction it is nearly zero, whereas in the perpendicular directions it takes finite values.
Therefore, anisotropy is prominent.
Moreover, due to weighting toward high-density regions, $\Delta K^{(N)}$ is larger than in voids, but does not increase isotropically as in halos.
On the other hand, $\mathcal{R}^{(N)}$ is effective as an indicator that removes the overall curvature amplitude and measures the relative importance of finite-dimensional effects, and provides auxiliary information to distinguish halos and filaments even when they have similar $\Delta K^{(N)}$.
\medskip

\noindent
This result is schematically depicted in Fig.~\ref{fig:local_curvature}.

\subsection{Construction of Classification Indicators}

For a set of directions $\{\mathbf{e}_a\}$,
\begin{align}
\overline{K}^{(\infty)}(x;R) &= \frac{1}{m}\sum_a K_{\mathrm{eff}}^{(\infty)}(x;R;\mathbf{e}_a), \\
\overline{K}^{(N)}(x;R) &= \frac{1}{m}\sum_a K_{\mathrm{eff}}^{(N)}(x;R;\mathbf{e}_a)
\end{align}
are defined as the mean curvature.
Furthermore, anisotropy is defined by
\begin{align}
A^{(\infty)}(x;R) &= \max_a K_{\mathrm{eff}}^{(\infty)}(x;R;\mathbf{e}_a)-\min_a K_{\mathrm{eff}}^{(\infty)}(x;R;\mathbf{e}_a),\\
A^{(N)}(x;R) &= \max_a K_{\mathrm{eff}}^{(N)}(x;R;\mathbf{e}_a)-\min_a K_{\mathrm{eff}}^{(N)}(x;R;\mathbf{e}_a)
\end{align}
Then, the direction-averaged difference and ratio are defined as
\begin{align}
\overline{\Delta K}^{(N)}(x;R)
&=
\overline{K}^{(N)}(x;R)-\overline{K}^{(\infty)}(x;R),\\
\overline{\mathcal{R}}^{(N)}(x;R)
&=
\frac{\overline{K}^{(N)}(x;R)}{\overline{K}^{(\infty)}(x;R)}
\end{align}
$\overline{\Delta K}^{(N)}$ represents the absolute increment of concentrated structures, and $\overline{\mathcal{R}}^{(N)}$ represents the relative importance of that increment.
However, when $\overline{K}^{(\infty)}(x;R)$ is close to zero, $\overline{\mathcal{R}}^{(N)}$ becomes unstable, and therefore it is appropriate to use it under a threshold condition in practice.

In this case, structures are characterized as follows:
\begin{itemize}
\item In voids, the mean curvature is small, anisotropy is also small, and $\overline{\Delta K}^{(N)}$ is also small.
\item In halos, the mean curvature is large, anisotropy is small, and both $\overline{\Delta K}^{(N)}$ and $\overline{\mathcal{R}}^{(N)}$ are large.
\item In filaments, the mean curvature is moderate, anisotropy is large, $\overline{\Delta K}^{(N)}$ is moderate, and $\overline{\mathcal{R}}^{(N)}$ serves as an auxiliary indicator of the relative strength of finite-dimensional effects.
\end{itemize}

When directional curvature is evaluated by the direction-dependent deformation
\begin{align}
T_{\varepsilon,\mathbf{e}}(y) = y+\varepsilon \langle y-x,\mathbf{e}\rangle \mathbf{e}
\end{align}
directional differences appear prominently in filaments.
Therefore, structure classification based on local effective curvature provides a theoretically consistent framework based on the second variation of entropy in Wasserstein geometry.

\section{Structure of Estimators as a Theoretical Consequence}
\label{sec:estimator_structure}

In this chapter, based on the local effective curvature introduced in the previous chapters, we clarify the structure of estimators that follows from this theory.
The purpose of the present study is not to provide a numerical implementation itself, but organizing what form of estimator is theoretically inevitable is indispensable for connecting this framework to finite data.
In particular, the estimator is to be understood not as an arbitrary statistical summary, but as an operation that extracts the entropy response to the deformation of local measures in Wasserstein geometry.

\subsection{Extraction Structure of Local Curvature}

The local effective curvature defined in Section~\ref{sec:local_curvature}, using the local measure $\mu_x^{(R)}$ and the reference measure $\nu_x^{(R)}$, essentially has the form
\begin{widetext}
\begin{align}
K_{\mathrm{eff}}(x;R)
=
\frac{2}{t(1-t)W_2^2(\mu_x^{(R)},\nu_x^{(R)})}
\left[
(1-t)\mathcal{E}(\mu_x^{(R)})
+
t\,\mathcal{E}(\nu_x^{(R)})
-
\mathcal{E}(\mu_{x,t}^{(R)})
\right]
\end{align}
\end{widetext}
In particular, in the present paper we adopted $t=1/2$ in order to preserve symmetry and the correspondence with the second variation.
Therefore, the construction of the estimator is decomposed into four elements: the construction of the local measure, the choice of the reference measure, the evaluation of the Wasserstein distance, and the evaluation of the entropy drop along the geodesic.

This means that the estimator in the present framework is defined not as a mere empirical fitting or an ad hoc statistic, but as a composite geometric operation for extracting geometric structure on measure space.
Indeed, the way of taking the local window is responsible for the scale of spatial coarse-graining, the way of choosing the reference measure is responsible for which deformation is regarded as the standard, the Wasserstein distance is responsible for the magnitude of the deformation, and the entropy difference is responsible for the response to that deformation.
Therefore, the degrees of freedom of the estimator are not arbitrary, and each of them corresponds to the geometric definitions in the previous chapters.

What is important here is that this structure agrees with the local expression of the entropy Hessian in the Otto formulation derived in Section~\ref{sec:structure_classification}.
That is, for the local linear deformation $v(y)=A(y-x)$,
\begin{align}
\tat{\mathcal{E}_{\infty}}{t}
\;\sim\;
\tr(H_x A\Sigma_x^{(R)}A^{\top})
\end{align}
holds, and therefore the estimator has a structure that directly evaluates the second variation of entropy at finite scale.
Accordingly, the present estimator is understood as a device for extracting the finite-scale version of Ricci curvature in Wasserstein geometry.

\subsection{Local Limit and Consistency}

In the local quadratic approximation,
\begin{align}
\ln \rho(y)
=
\ln \rho(x)
-
\frac{1}{2}(y-x)^{\top}H_x(y-x)
+
O(|y-x|^3)
\end{align}
can be written.
In this case, the local effective curvature is
\begin{align}
K_{\mathrm{eff}}^{(\infty)}(x;R;A)
=
\frac{\tr(H_x A\Sigma_x^{(R)}A^{\top})}
{\tr(A\Sigma_x^{(R)}A^{\top})}
+
O(R)
\end{align}
and therefore reproduces the directional information of the local log-density curvature $H_x$ in the limit $R\to0$.
That is, the present framework gives the curvature structure of the log-density in the sense of the local quadratic approximation for continuous density fields, and at finite scale extracts its coarse-grained geometric information.

In this sense, the estimator has a consistent scale structure in which it converges to the local structure in the limit $R\to0$, and at finite $R$ returns an effective geometric quantity corresponding to the observational resolution.
Therefore, in the present theory an estimator should be understood not as a device that directly estimates a differential quantity at a single point, but as a device that evaluates the response of measure deformation stabilized through a local window.
This point means, also when giving a numerical implementation later, that the choice of the smoothing radius and the local window is not a mere technical detail, but part of the definition of the geometric quantity itself.

Furthermore, this limiting structure is directly connected with the classification indicators introduced in Section~\ref{sec:structure_classification}.
That is,
\begin{align}
\overline{K}^{(\infty)} \;\longrightarrow\; \frac{1}{3}\tr H_x,
\qquad
A^{(\infty)} \;\longrightarrow\; \lambda_{\max}-\lambda_{\min}
\end{align}
holds, and the estimator gives a finite-scale version of the local eigenvalue structure.

\subsection{Extension to Finite Samples}

Since observational data are given as finite sets of points, the local measure $\mu_x^{(R)}$ is constructed as a discrete measure.
In this case, the Wasserstein distance is defined as a discrete optimal transport problem, and the reference measure $\nu_x^{(R)}$ is likewise given as a discrete or smoothed reference measure.
Therefore, the framework constructed in this paper can move from the theory for continuous measures to the theory for finite point processes while preserving the structure.

What is important here is that what changes by finite sampling is primarily the representation of the measure, whereas the logical structure of the estimator itself does not change.
That is, the estimator is still composed of the setting of the local window, the construction of the reference measure, the computation of discrete optimal transport, and the evaluation of the entropy functional.

Furthermore, the direction average, anisotropy, difference $\Delta K^{(N)}$, and ratio $\mathcal{R}^{(N)}$ introduced in Section~\ref{sec:structure_classification} are positioned as secondary indicators that integrate these basic local curvature estimates and connect them to cosmological structure classification.
In particular, $\Delta K^{(N)}$ measures the absolute strength of high-density concentration, and $\mathcal{R}^{(N)}$ functions as a relative indicator that measures to what extent that concentration effect is dominant with respect to the total curvature, and therefore the two carry essentially different information.
Therefore,
\begin{align}
\begin{array}{c}
\text{local effective curvature in a continuous density field} \\[4pt]
\big\downarrow \\[4pt]
\text{local curvature estimation on finite samples} \\[4pt]
\big\downarrow \\[4pt]
\text{aggregation into classification indicators}
\end{array}
\end{align}
this hierarchical structure itself is the essence of the estimator that follows from the present theory.
The discussion in this chapter does not provide the details of implementation, but it makes explicit from what geometric components the estimators to be adopted in future numerical and observational studies should be composed.

\section{Discussion and Conclusions}
\label{sec:discussion_conclusion}

\subsection{Local limit and Hessian structure}
\label{subsec:local_limit_Hessian}

In this paper, we have constructed a geometric framework for cosmological large-scale structure based on optimal transport theory and Wasserstein geometry, in which curvature on the probability measure space $\mathcal{P}_2(M)$ is defined through the geodesic convexity of entropy.
Within this formulation, the effective Ricci curvatures $K_{\mathrm{eff}}^{(\infty)}$ and $K_{\mathrm{eff}}^{(N)}$ provide a unified description of infinite-dimensional diffusive geometry and finite-dimensional concentration effects.
Furthermore, curvature is realized as the second variation of entropy in the geodesic direction under the Otto metric, establishing a direct connection between geometric structure and variational principles on measure space.

By introducing local and reference measures, we have localized this curvature to finite scales and formulated observable quantities.
Under the local quadratic approximation, the local effective curvature reduces to the curvature matrix of the log-density,
\begin{align}
H_x=-\nabla^2\ln\rho(x),
\end{align}
and its eigenvalue structure determines local geometry.
In this limit, conventional Hessian-type classifications of cosmic structures are naturally recovered, showing that existing classification schemes arise as limiting cases of the present framework.
In particular, mean curvature and anisotropy correspond respectively to the trace and eigenvalue dispersion of $H_x$, providing a geometric reinterpretation of structure classification within Wasserstein geometry.

\subsection{Scale-dependent curvature and multiscale structure}
\label{subsec:scale_dependent_curvature}

A central feature of the present framework is that curvature depends explicitly on the scale $R$.
This dependence reflects not a simple smoothing operation, but a redefinition of local measures, and is therefore interpreted as a scale flow on measure space.
Unlike Ricci flow, this does not describe temporal evolution of geometry, but rather a change of resolution for a fixed geometry.

As a result, curvature and anisotropy may exhibit non-monotonic behavior and even sign changes due to the mixing of different structures.
This implies that structure classification is intrinsically scale-dependent.
From this viewpoint, cosmological structures are characterized not by fixed geometric types, but by their response to coarse-graining and their trajectories in scale space.

\subsection{Toward curvature statistics and observables}
\label{subsec:curvature_statistics}

By treating the local effective curvature as a random field, the present framework naturally leads to a statistical description in terms of curvature.
Under the local quadratic approximation, curvature statistics are expressed as weighted integrals of the power spectrum and derivatives of the correlation function, corresponding to $k^2 P(k)$ and $k^4 P(k)$ contributions in Fourier space.
Therefore, curvature statistics emphasize geometric structure rather than amplitude fluctuations, and provide information complementary to conventional two-point statistics.

In particular, R\'{e}nyi-based indicators such as $\Delta K^{(N)}$ and $\mathcal{R}^{(N)}$ enhance sensitivity to high-density and nonlinear structures.
This suggests that curvature-based statistics can provide independent constraints on structure formation beyond those accessible through standard correlation functions and power spectra.

The present framework also imposes structural constraints on estimators for finite samples.
Local curvature estimation is constructed from geometric operations consisting of local measure definition, reference measure selection, Wasserstein distance evaluation, and entropy variation.
On this basis, mean curvature, anisotropy, and R\'{e}nyi corrections are combined to yield classification indicators.
Thus, the present theory provides a bridge between geometric formulations for continuous density fields and practical estimators for discrete observational data.

\subsection{Outlook}

Several important directions naturally arise from the present framework.
These include the construction and validation of numerical estimators, applications to cosmological simulations and observational data, and a systematic analysis of the independent information carried by curvature statistics.
In particular, the response-based nature of curvature suggests a deeper connection with response theory and symmetry-based formulations, which may further clarify the physical interpretation of the present framework.

Finally, the present framework suggests a natural geometric reinterpretation of peak statistics in cosmology.
Since local effective curvature reduces to the Hessian of the log-density in the local limit, peak-based descriptions such as the BBKS formalism can be understood as limiting cases of curvature-induced structures in measure space.
This perspective indicates that peak statistics may be systematically reconstructed as curvature-induced structures within measure space. 
A detailed formulation of this connection will be presented in future work.

\section*{Acknowledgments}

I am deeply grateful to Shiro Ikeda for helpful insights for this work.
This work was supported by JSPS Grant-in-Aid for Scientific Research (24H00247) and by the Joint Research Program of the Institute of Statistical Mathematics (General Research 2), ``Machine-learning cosmogony: from structure formation to galaxy evolution.''

\bibliographystyle{apsrev4-2}
\bibliography{curvature_statistics_formulation}

\appendix

\section{Derivation of the Volume Distortion Coefficient $\beta_t^{K,N}$ by Jacobi Fields}
\label{app:beta_derivation}

In this appendix, we briefly summarize the geometric origin of the volume distortion coefficient $\beta_t^{K,N}$ used in Section~\ref{subsec:renyi_cdKN}.
This coefficient is derived by comparing the deformation of volume elements along geodesics under a lower bound on Ricci curvature, and becomes a fundamental control factor of density deformation in the curvature-dimension condition CD$(K,N)$ \citep{Ohta2010RicciEntropyOT}.
The derivation uses the method of evaluating volume deformation from the comparison of Jacobi fields, and from this obtaining the volume distortion coefficient.

\subsection{Jacobi Fields and Volume Elements}

Consider a geodesic $\gamma(t)$ on a Riemannian manifold $(M,g)$.
A Jacobi field $J(t)$ along the geodesic satisfies
\begin{align}
    \frac{{\rm D}^2J}{{\rm d}t^2}+R(J,\dot{\gamma})\dot{\gamma}=0
\end{align}
using the covariant derivative ${\rm D}/{\rm d}t$ with respect to the Levi-Civita connection.
Taking $(N-1)$ Jacobi fields $J_1,\dots,J_{N-1}$ orthogonal to the geodesic, the volume element spanned by them is given by
\begin{align}
\mathcal{J}(t)=\det\bigl(J_1(t),\dots,J_{N-1}(t)\bigr)
\end{align}
This $\mathcal{J}(t)$ represents the volume deformation along the geodesic.

\subsection{Control of Volume Deformation by Ricci Curvature}

Ricci curvature controls the average behavior of Jacobi fields and imposes constraints on the deformation of volume elements.
Formally, one obtains an estimate
\begin{align}
\pat{}{t}\ln \mathcal{J}(t)
\le
-\mathrm{Ric}(\dot{\gamma},\dot{\gamma})
\end{align}
In particular, assuming the lower bound of Ricci curvature
\begin{align}
\mathrm{Ric}\ge K
\end{align}
one has
\begin{align}
\pat{}{t}\ln \mathcal{J}(t)\le -K
\end{align}
and the contraction and expansion of the volume element along the geodesic are uniformly controlled.

\subsection{Comparison with Model Spaces}

In order to compare this inequality, consider a model space with constant curvature $K$.
In this case, the magnitude of the Jacobi field is given by
\begin{align}
s_K(r)
=
\begin{cases}
\dfrac{1}{\sqrt{K}}\sin(\sqrt{K}r), & K>0,\\
r, & K=0,\\
\dfrac{1}{\sqrt{-K}}\sinh(\sqrt{-K}r), & K<0
\end{cases}
\end{align}
Therefore, the corresponding volume element is
\begin{align}
\mathcal{J}_{\mathrm{model}}(r)=s_K(r)^{N-1}
\end{align}

\subsection{Definition of the Volume Distortion Coefficient}

For the geodesic length $r=d(x,y)$, define the volume distortion coefficient from the comparison of volume deformation at time $t$ by
\begin{align}
\beta_t^{K,N}(r)
=
\frac{\mathcal{J}_{\mathrm{model}}(tr)}{t^{N-1}\mathcal{J}_{\mathrm{model}}(r)}
\end{align}
From this one obtains
\begin{align}
\beta_t^{K,N}(r)
=
\left[
\frac{s_K(tr)}{t\,s_K(r)}
\right]^{N-1}
\end{align}

\subsection{Explicit Form in the Positive-Curvature Case}

In particular, when $K>0$,
\begin{align}
s_K(r)=\frac{1}{\sqrt{K}}\sin(\sqrt{K}r)
\end{align}
and therefore
\begin{align}
\beta_t^{K,N}(r)
=
\left[
\frac{\sin\bigl(t r\sqrt{K}\bigr)}
{t\,\sin\bigl(r\sqrt{K}\bigr)}
\right]^{N-1}
\end{align}
Furthermore, making the dimension $N$ explicit, one can write
\begin{align}
\beta_t^{K,N}(r)
=
\left[
\frac{\sin\bigl(t r\sqrt{K/(N-1)}\bigr)}
{t\,\sin\bigl(r\sqrt{K/(N-1)}\bigr)}
\right]^{N-1}
\end{align}

\subsection{Relation to Optimal Transport}

Since the Jacobian of the optimal transport map $\Phi_t$ is given by the determinant of Jacobi fields,
\begin{align}
\det D\Phi_t \gtrsim \beta_t^{K,N}(r)
\end{align}
one obtains a comparison estimate.
By combining this with the density transformation formula
\begin{align}
\rho_t(\Phi_t(x)) = \frac{\rho_0(x)}{\det D\Phi_t(x)}
\end{align}
one obtains the density inequality of the CD$(K,N)$ condition stated in Section~\ref{subsec:renyi_cdKN}.

The volume distortion coefficient $\beta_t^{K,N}$ is a quantity that makes explicit, by comparison with model spaces, the behavior of Jacobi fields controlled by Ricci curvature.
This coefficient is a fundamental control factor of density deformation in optimal transport, and provides the geometric core of the curvature-dimension condition.

\section{Correspondence Between Local Effective Curvature and Hessian Structure}
\label{app:local_limit}

In this appendix, we formulate the correspondence between the local effective curvature described in Section~\ref{subsec:local_limit_Hessian} and the curvature matrix of the log-density $H_x=-\nabla^2\ln\rho(x)$ under the local quadratic approximation.
In particular, we show that the directional curvature is given as a Rayleigh quotient, and that the mean curvature and anisotropy correspond to the eigenvalue structure of $H_x$.

When the density distribution is sufficiently smooth, in the neighborhood of a point $x$ one can expand
\begin{align}
\ln \rho(y)
=
\ln \rho(x)
-
\frac{1}{2}(y-x)^{\top}H_x(y-x)
+
O(|y-x|^3)
\end{align}
Here $H_x=-\nabla^2\ln\rho(x)$ is a symmetric matrix that describes the local geometric structure.
Under this approximation, the local measure $\mu_x^{(R)}$ is approximated by a Gaussian distribution, and its covariance matrix can be written as
\begin{align}
\Sigma_x^{(R)}
=
c R^2 I
+
O(R^4)
\end{align}

The local effective curvature along a direction $\mathbf{e}$ is given by
\begin{align}
K_{\mathrm{eff}}^{(\infty)}(x;R;\mathbf{e})
=
\frac{\mathbf{e}^{\top}H_x \Sigma_x^{(R)} \mathbf{e}}
{\mathbf{e}^{\top}\Sigma_x^{(R)} \mathbf{e}}
+
O(R)
\end{align}
In the limit where the covariance is isotropic, this expression reduces to
\begin{align}
K_{\mathrm{eff}}^{(\infty)}(x;R;\mathbf{e})
=
\mathbf{e}^{\top}H_x\mathbf{e}
+
O(R)
\end{align}
Therefore, the local effective curvature is interpreted as the Rayleigh quotient with respect to the matrix $H_x$.
In particular, along eigenvector directions, the corresponding eigenvalues appear directly as curvature.

Taking the directional average of $K_{\mathrm{eff}}^{(\infty)}(x;R;\mathbf{e})$, by averaging over the unit sphere one obtains
\begin{align}
    \overline{K}^{(\infty)}(x;R) = \frac{\tr H_x}{d}+ O(R) \label{eq:average_Keff}
\end{align}
Therefore, the mean curvature coincides with the trace of $H_x$.
Moreover, the directional variance of curvature is defined as $\vari \bigl(K_{\mathrm{eff}}^{(\infty)}(x;R;\mathbf{e})\bigr)$, and this corresponds to the variance of the eigenvalues of $H_x$.
That is, when the eigenvalues are equal, the curvature becomes isotropic, and when there is dispersion among the eigenvalues, anisotropy appears.

Let the eigenvalues of $H_x$ be $\lambda_1,\dots,\lambda_d$.
Then the sign structure of the local effective curvature is determined by these signs:
\begin{itemize}
    \item When all eigenvalues are negative, the curvature is negative in all directions, and the distribution is expansive.
    This corresponds to a void structure.
    \item When all eigenvalues are positive, the curvature is positive in all directions, and the distribution is contractive.
    This corresponds to a halo structure.
    \item When only some of the eigenvalues are positive, the curvature differs depending on direction, and anisotropy becomes prominent.
    In this case, filament or sheet structures appear depending on the number of positive eigenvalues.
\end{itemize}
In this way, the local effective curvature completely characterizes local geometry through the eigenvalue structure of $H_x$.
Therefore, conventional Hessian-type classification is naturally reproduced in the present framework.
This result shows that the local effective curvature coincides with the second-derivative structure of the log-density.

\section{Asymptotic Structure and Mixing Effects of Scale-dependent Curvature}
\label{app:scale_flow_derivation}

In this appendix, we formulate the behavior of scale-dependent curvature introduced in Section~\ref{subsec:scale_dependent_curvature} based on the local quadratic approximation and asymptotic expansion.
In particular, we derive the scale expansion of the local covariance, and the conditions for non-monotonicity and sign change of curvature due to mixing of multiple structures.

\subsection{Scale Expansion of Local Covariance}

Consider the covariance matrix of the local measure $\mu_x^{(R)}$
\begin{align}
\Sigma_x^{(R)}
=
\int_{B_R(x)} (y-x)(y-x)^{\top}\,\pd\mu_x^{(R)}(y)
\end{align}
Assuming that the density is sufficiently smooth and that local isotropy is dominant, by symmetry $\Sigma_x^{(R)}$ is expanded as an even function of the scale $R$.
That is,
\begin{align}
\Sigma_x^{(R)}
=
c_2 R^2 I
+
c_4 R^4 M_x
+
O(R^6)
\end{align}
Here $c_2, c_4$ are constants, and $M_x$ is a symmetric matrix determined by the fourth-order moment of the density.
The first term represents isotropic spread, and the second term carries anisotropy and the influence of local structure.

\subsection{Scale Expansion of Curvature}

The local effective curvature is given by
\begin{align}
K_{\mathrm{eff}}^{(\infty)}(x;R;A)
=
\frac{\tr(H_x A\Sigma_x^{(R)}A^{\top})}
{\tr(A\Sigma_x^{(R)}A^{\top})}
\end{align}
Substituting the expansion of the covariance into this expression, the numerator and denominator have terms of order $R^2$ and $R^4$, respectively.
Arranging these, one obtains
\begin{align}
K_{\mathrm{eff}}^{(\infty)}(x;R;A)
=
\frac{\tr(H_x A)}{\tr(A)}
+
R^2 \cdot \mathcal{C}_x(A)
+
O(R^3)
\end{align}
Here $\mathcal{C}_x(A)$ is a correction term depending on $M_x$ and $H_x$, and represents the contribution of anisotropy and higher-order structure.
From this result, it is seen that the scale dependence of curvature is controlled by higher-order moments and is in general non-monotonic.

\subsection{Non-monotonicity and Structural Transition in a Mixing Model}

Next, consider the case where different geometric structures coexist.
Approximate the local measure as
\begin{align}
\mu_x^{(R)}
\simeq
w(R)\,\mu_{\mathrm{halo}}
+
\bigl(1-w(R)\bigr)\,\mu_{\mathrm{fil}}
\end{align}
Here $w(R)$ is a scale-dependent weight, which generally decreases as $R$ increases.
In this case, the effective covariance and curvature are approximated as weighted averages of the contributions of each structure.
Therefore, the curvature can be written as
\begin{align}
K_{\mathrm{eff}}(R)
\simeq
w(R)\,K_{\mathrm{halo}}
+
\bigl(1-w(R)\bigr)\,K_{\mathrm{fil}}
+
\text{correction terms}
\end{align}
From this expression, it is seen that curvature changes non-monotonically as $w(R)$ varies.
In particular, when the signs or magnitudes of $K_{\mathrm{halo}}$ and $K_{\mathrm{fil}}$ differ, extrema or inflection points of curvature appear.

The sign of curvature is directly related to structure classification.
In the above mixing model, the condition under which the effective curvature becomes zero is given by
\begin{align}
w(R_\ast)\,K_{\mathrm{halo}}
+
\bigl(1-w(R_\ast)\bigr)\,K_{\mathrm{fil}}
=0
\end{align}
At the scale $R_\ast$ satisfying this condition, the sign of curvature reverses.
This sign reversal implies that the classification of local geometric structure changes depending on scale.
That is, a region identified as a halo at small scale may appear as a filamentary structure at larger scale.
Therefore, the scale dependence of local effective curvature functions as a quantity that characterizes structural boundaries and transitions.
In this sense, the scale flow provides a continuous extension of structure classification.

The above results show that scale-dependent curvature is naturally derived from the local quadratic approximation and mixing effects.

\section{Correlation Functions and Power Spectrum from Curvature Statistics}
\label{app:curvature_statistics}

In this appendix, we formulate the fundamental structure of curvature statistics introduced in Section~\ref{subsec:curvature_statistics} in terms of their relation to the correlation function and power spectrum of the density field.
In particular, we treat the local effective curvature as a random field and make explicit that its statistical quantities correspond to the differential structure of the density field.
Following the convention in cosmology, introduce the density contrast
\begin{align}
    \delta(x)=\dfrac{\rho(x)}{\bar{\rho}} - 1
\end{align}
and assume that $\delta$ is a Gaussian random field with zero mean.
Since the local effective curvature $K_{\mathrm{eff}}(x;R)$ is defined as a function of the density field $\rho(x)$, it becomes a random field for a stochastic density field.
Under the local quadratic approximation,
\begin{align}
\overline{K}^{(\infty)}(x;R)
=
-\frac{1}{d}\,\nabla^2 \ln \rho_R(x)
+
O(R)
\end{align}
can be written.
Here $\rho_R$ is the density smoothed at scale $R$.
Therefore, the curvature field is governed by the Laplacian of the log-density.
In Fourier space,
\begin{align}
    \delta(x) &= \int \tilde{\delta}(k)\,e^{i k\cdot x}\,\pd k, \\
    \expt\bigl[\tilde{\delta}(k)\tilde{\delta}(k')\bigr] &= (2\pi)^d \deltadir (k+k') P(k)
\end{align}
hold.

The smoothed density can be written using the window function $W_R(k)$ as
\begin{align}
\delta_R(x)
=
\int \tilde{\delta}(k) W_R(k) e^{i k\cdot x}\,\pd k
\end{align}
Then
\begin{align}
\nabla^2 \delta_R(x)
=
-\int k^2 \tilde{\delta}(k) W_R(k) e^{i k\cdot x}\,\pd k
\end{align}
Therefore, the expectation value of curvature is
\begin{align}
\expt \bigl[\overline{K}^{(\infty)}(x;R)\bigr]
=
\frac{1}{d}
\int k^2 P(k) W_R(k)\,\pd k
+
O(R)
\end{align}
Similarly, the variance is
\begin{align}
\vari \bigl[\overline{K}^{(\infty)}(x;R)\bigr]
=
\frac{1}{d^2}
\int k^4 P(k) W_R(k)^2\,\pd k
\end{align}
From this result, it is seen that curvature statistics assign $k^2$ and $k^4$ weights to the power spectrum.
Therefore, curvature has strong sensitivity to high-frequency components and small-scale structures of the density field.

Next, consider the two-point correlation of the curvature field.
For distance $r=|x-y|$,
\begin{align}
\cov \bigl(K_{\mathrm{eff}}(x;R),K_{\mathrm{eff}}(y;R)\bigr) = \frac{1}{d^2}\,\nabla^2 \nabla^2 \xi_R(r)
\end{align}
holds.
Here $\xi_R$ is the smoothed correlation function.
This expression shows that curvature correlation is given as the action of the Laplacian twice on the correlation function.
Therefore, curvature statistics are interpreted as statistical quantities corresponding to spatial derivatives of the correlation function.

Furthermore, considering the R\'{e}nyi-type correction, smoothing based on the weighted density
\begin{align}
\rho^{1-\frac{1}{N}}(x)
\end{align}
is introduced.
In this case, curvature can be written as
\begin{align}
K_{\mathrm{eff}}^{(N)}(x;R)
=
-\frac{1}{d}\,\nabla^2 \ln \rho_R^{(N)}(x)
+
\text{correction terms}
\end{align}
Here $\rho_R^{(N)}$ is the effective density with R\'{e}nyi weighting.
Since this weighting enhances the contribution of high-density regions,
\begin{align}
\Delta K^{(N)}(x;R)
=
K_{\mathrm{eff}}^{(N)}(x;R)
-
K_{\mathrm{eff}}^{(\infty)}(x;R)
\end{align}
functions as an indicator of nonlinear structures.

From the above, curvature statistics correspond to the statistical structure of the density field through the transformation
\begin{align}
P(k)
\;\longrightarrow\;
k^2 P(k), \quad k^4 P(k)
\end{align}
and in real space,
\begin{align}
\xi(r)
\;\longrightarrow\;
\nabla^2 \nabla^2 \xi(r)
\end{align}
Therefore, curvature statistics are understood as operators that map amplitude statistics of the density field to geometric statistics.
In this sense, although curvature statistics are not completely independent of conventional correlation functions or power spectra, they provide complementary information with different scale sensitivity through spatial derivatives.

\end{document}